\documentclass[amsmath, amssymb, superscriptaddress, reprint, twocolumn, colorlinks=true, citecolor=blue, linkcolor=blue, urlcolor=blue]{revtex4-1}
\pdfoutput=1
\usepackage{natbib}
\usepackage[utf8]{inputenc}
\usepackage[T1]{fontenc}
\usepackage{graphicx}
\usepackage{grffile}
\usepackage{longtable}
\usepackage{wrapfig}
\usepackage{rotating}
\usepackage[normalem]{ulem}
\usepackage{amsmath}
\usepackage{textcomp}
\usepackage{amssymb}
\usepackage{capt-of}
\usepackage{hyperref}
\usepackage{mathptmx}
\usepackage{upgreek}
\usepackage{float}
\usepackage{bm}
\usepackage{setspace}
\usepackage{booktabs}
\usepackage{subfigure}
\usepackage{url}
\usepackage{titlesec}
\date{\today; Authors to whom correspondence should be addressed: fuyangyang@tsinghua.edu.cn}
\title{}

\begin{document}

\title{Microscopic characteristics of SF\(_6\) partial discharge induced by a floating linear metal particle}

\author{Zihao Feng}
\affiliation{Department of Electrical Engineering, Tsinghua University, Beijing 100084, China}
\author{Yuanyuan Jiang}
\affiliation{Department of Electrical Engineering, Tsinghua University, Beijing 100084, China}
\author{Liyang Zhang}
\affiliation{Department of Electrical Engineering, Tsinghua University, Beijing 100084, China}
\author{Zhigang Liu}
\affiliation{Department of Electrical Engineering, Tsinghua University, Beijing 100084, China}
\author{Kai Wang}
\affiliation{Tsinghua Shenzhen International Graduate School, Shengzhen, Guangdong 518055, China}
\author{Xinxin Wang}
\affiliation{Department of Electrical Engineering, Tsinghua University, Beijing 100084, China}
\author{Xiaobing Zou}
\affiliation{Department of Electrical Engineering, Tsinghua University, Beijing 100084, China}
\author{Haiyun Luo}
\affiliation{Department of Electrical Engineering, Tsinghua University, Beijing 100084, China}
\author{Yangyang Fu*}
\affiliation{Department of Electrical Engineering, Tsinghua University, Beijing 100084, China}
\affiliation{State Key Laboratory of Power System Operation and Control, Department of Electrical Engineering, Tsinghua University, Beijing 100084, China}
\begin{abstract}
Direct current (DC) gas insulated transmission lines (GILs) have been widely used in power transmission, but might be threatened by partial discharge due to the presence of floating impurities (e.g., dust and metal particles) inside the sealed chamber. In this letter, by using a 2D fluid model we characterize the microscopic properties of the partial discharge induced by a floating linear metal particle in SF\(_6\) (both the discharge propagation and interaction between space charge and metal particle) under negative high voltage direct current (HVDC) conditions. Due to the strong electronegativity of SF\(_6\), the spatiotemporal distributions of the charged species (electrons, positive and negative ions), space charge, and reduced electric field are rather different from those in air. Notably, a negative ion region is observed around the top tip of the metal particle, and it plays an important role in the generation and propagation of primary and secondary streamers in SF\(_6\), which may lead to severe motion characteristics of the particle and aliasing of partial discharge signals. Additionally, we analyze the charging process and electric force reversal phenomenon, which may provide a more precise understanding of the underlying mechanisms of the firefly motion previously reported for DC GILs.
\end{abstract}

\maketitle
Direct current (DC) gas insulated transmission lines (GILs) possess unique advantages in urban underground power transmission and underwater power transmission applications. In practice, SF\(_6\) partial discharge is usually induced due to the presence of metal particles, and existing particle detection and suppression measures have been proven ineffective for DC GILs \cite{10007898,trap}. Although a considerable amount of experimental results have indicated that the complicated partial discharge induced by metal particles is the main cause of the above issue\cite{exp3,exp4}, the microscopic characteristics of related discharge processes and the space charge effect on particles remain unclear.

Gas discharge simulations can resolve spatiotemporal microscopic characteristics of the discharge \cite{26}. Chen \textit{et al.} \cite{10.1063/5.0104205}simulated metal particle-induced breakdown within a 200 $\upmu$m microgap. Sun \textit{et al.} \cite{Sun_2021}and Zhong \textit{et al.} \cite{https://doi.org/10.1002/ctpp.202100133}simulated the impact of the metal particle's field electron emission on breakdown. However, the discharge gas used in their models is different from that used in GILs. To date, simulations of SF\(_6\) have been reported primarily for single-gap electrode configurations\cite{SF6111,sf6333,14,sf61991,SF62002,15,19}, and relating fluid models have typically utilized the local field approximation, which may introduce inaccuracies in capturing non-local characteristics, particularly, the dynamics of charged species near the computational boundary. Moreover, models without relying on the particle-structure scenario cannot obtain the multistage discharge process and interaction between space charge and metal particles, compared to combined gas gap with floating particles in GILs. In particular, under the negative high voltage, linear metal particles often exhibit a firefly motion and significantly reduce the efficiency of diagnosis and suppression \cite{Zhuang_2023,4074282}. One key determinant of firefly motion is the reversal of the electric field force affected by space charge. Chang \textit{et al.} \cite{Chang_2023} used the point charge model to analyze this force and reported that the polarity of the overall charge $Q$ of a metal particle reverses when firefly motion occurs. However, the surface charge density of linear particles is highly nonuniform and directly determined by the transient interaction between space charge and metal particle, which should be further investigated. To date, a precise understanding of the microscopic characteristics of SF\(_6\) discharge induced by floating metal particles is still awaiting.

\begin{figure}[h]
\centering
\includegraphics[clip, trim=0cm 0cm 0cm 0cm, width=8.5cm]{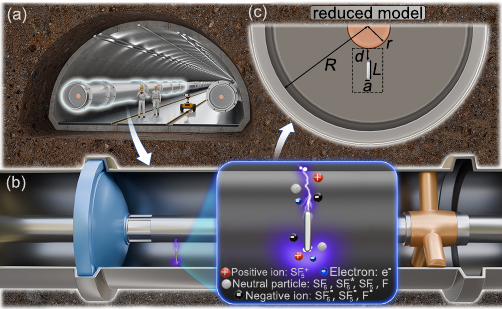}
\caption{\label{Fig.1} Schematic of (a) the GILs in a high-voltage power transmission system, (b) the partial discharge induced by a floating linear metal in SF\(_6\) under HVDC conditions, and (c) the reduced model of the coaxial electrodes with \(R\) and \(r\) as radii, \(L\), \(a\), \(d\) as the length, width and distance between the particle top tip and the central electrode.}
\end{figure}    

In this letter, we report the microscopic discharge characteristics (both the discharge propagation and interaction between space charge and metal particle) of SF\(_6\) partial discharge induced by a linear metal particle in a coaxial cylinder electrode and reveal the physical mechanism of the metal particle's electric field force reversal phenomenon, which usually occurs under negative DC high voltage and largely affects the firefly motion. A 2D fluid simulation model using local energy approximation is established, including plasma fluid equations \cite{equation} and 18 dominant plasma chemical reactions \cite{16,reaction2,reaction3}. 
\begin{figure}[b]
\centering
\includegraphics[clip,trim=0cm 0cm 0cm 0cm,width=8.5cm]{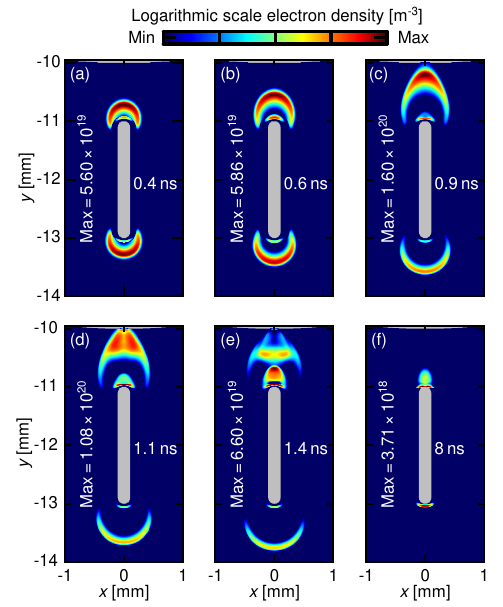}
\caption{\label{Fig.2}Spatiotemporal evolution of electron density during double-headed streamer propagation, (a)--(f) correspond to 0.4 ns, 0.6 ns, 0.9 ns, 1.1 ns, 1.4 ns, and 8 ns, respectively.}
\label{Fig.2}
\end{figure}
The critical reduced electric field \((E/N)_{\text{cr}}\) for SF\(_6\) effective ionization calculated by BOLSIG+ \cite{bosig+} is approximately 360 Td, which is in good agreement with the measured results of Christophorou \textit{et al} \cite{EN3601} and Morrow \cite{EN3602}, validating the rationality of the reaction system. For the boundary conditions for the floating metal particle, the current continuity equation \({\partial \sigma_{\text{s}}}/{\partial t}=\mathbf{n} \cdot \mathbf{J}_i+\mathbf{n} \cdot \mathbf{J}_e\) is used to represent the effect of plasma on particle charge, where \(\sigma_\text{s}\) is the surface charge density and \(\mathbf{n} \cdot \mathbf{J}_i\) and \(\mathbf{n} \cdot \mathbf{J}_e\) represent the normal components of the total ion current density and the total electron current density on the particle surface, respectively. The equipotential condition \(V \equiv \text { constant}\) is set for the metal particle surface, but time-dependant, where \(V\) is the floating potential of the particle. Then, an integral boundary condition is set to control the overall charge \(Q\) of the metal particle: 
\begin{equation}
   \label{Eq3}
\int_{\text{S}} \mathbf{n}\cdot\mathbf{D} 
 {\text{dS}}=Q
\end{equation}
where \(\mathbf{n}\cdot\mathbf{D}\) represents the normal components of the electric displacement on the particle surface. The above settings ensure that the electric field on the metal particle surface is normal to the surface and that the entire charge on the metal particle is distributed on the surface. The distribution of surface charge satisfies the electrostatic induction conditions of the metal, which ensures that the metal particle is in a state of electrostatic equilibrium at all time. The proposed model could reveal detailed mechanisms for the discharge induced by the floating metal particle. By employing the local energy approximation, the accuracy of the microscopic characteristics, particularly the dynamics of charged species near particle tips, can be ensured. The approach and findings presented in this letter could contribute to existing research.

A schematic of the discharge system is shown in Fig.\ref{Fig.1}. Here for simplicity, a 2D model is established to describe the circular cross-section of the coaxial cylinder electrode. The 2D description generally underestimates the local field enhancement (the field strength around protrusion), which thus cannot quantitatively capture the 3D characteristics of the partial discharge (e.g., transport of the charged species) but can provide a qualitative prediction of the general discharge mechanisms. For simplicity, we use a reduced model similar to that used in many experimental studies \cite{Chang_2023,expscaled2,expscaled3,particularregion}, the underlying physical mechanisms captured by the reduced model are consistent with those derived from the normal-sized model. The inner radius $r$ is 10 mm and the outer radius $R$ is 30 mm. The radius ratio $r/R$ is close to the optimum value of 1/2.718 \cite{radio}, which reduces the nonuniformity of the electric field distribution. The linear particle length $L$ is 2 mm, and the particle diameter $a$ is 0.2 mm. 

\begin{figure}[t]
\centering
\includegraphics[clip, trim = 0cm 0cm 0cm 0cm, width=8.5cm]{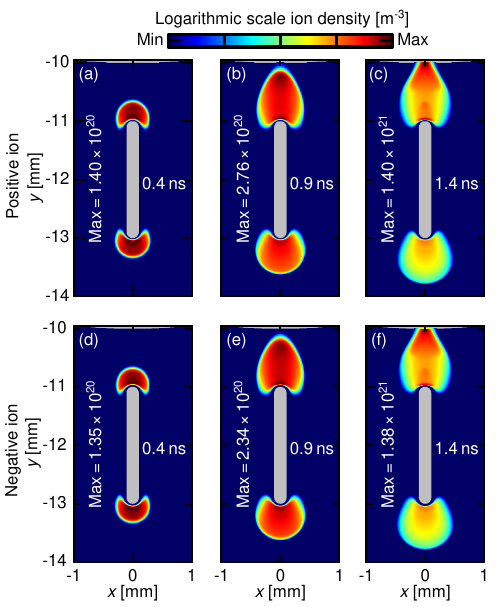}
\caption{\label{Fig.3}Spatiotemporal evolutions of positive ion density and negative ion density during the double-headed streamer propagation at 0.4 ns, 0.9 ns, and 1.4 ns, respectively. (a)--(c) correspond to positive ion density, (d)--(f) correspond to negative ion density.}
\end{figure}
\begin{figure*}[t]
\centering
\includegraphics[width=17cm]{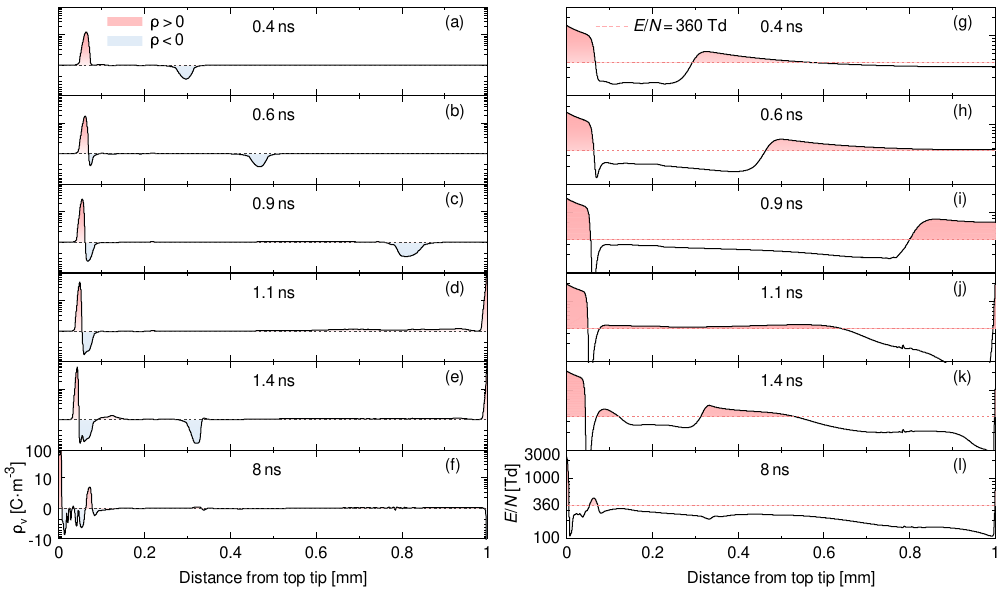}
\caption{\label{Fig.4}Spatial distributions of (a)--(f) the space charge \(\rho_\text{v}\) [unit in \(\text{C} \cdot \text{m}^{-3}\)] and (g)--(l) the reduced electric field \(E/N\) [unit in Td, 1 \(\text{Td} = 10^{-21}\) \(\text{V} \cdot \text{m}^{2}\)] along the axis from the top tip toward the high voltage electrode. The critical value \(E/N = 360\) Td corresponds to the breakdown threshold of SF\(_6\) gas.}
\end{figure*}
Since the timescale of discharge ($\sim$ ns) is much shorter than that of particle movement ($\sim$ ms), the particle is assumed to be floating in the gap. It locates near the central electrode and the entire gap is divided into two parts, with an upper gap distance \textit{d} of 1 mm between the particle top tip and central electrode. This distance is close to the experimental condition of firefly motion reported in Ref.~\cite{8684218}. The gas pressure \textit{p} is set to atmospheric, and thus this $pd$ range qualitatively ensures the same streamer discharge process as that of higher pressure \cite{pd}. According to existing experimental results \cite{7909196,21,23,25}, it can be inferred that -25 kV is between the particle lifting voltage and the breakdown voltage. As a result, the central electrode is set as a high voltage direct current (HVDC) electrode at a constant -25 kV, and the initial charge of particle $Q_{\text{0}}$ is set to -113.65 pC, which is calculated using formula \cite{Q01}:
\begin{equation}
   \label{Eq4}
Q_0=\frac{\pi  \varepsilon  L^2  E_r}{\ln \left(\frac{2 L}{a}\right)-1}
\end{equation}
where $\varepsilon$ is the permittivity of SF\(_6\), and $E_r$ is the electric field on the HVDC electrode surface. These settings ensure that the simulation represents the condition when the metal particle moves near the HVDC electrode after colliding with it.

The electric field is calculated self-consistently based on the Poisson's equation, and the charge transport is described by the fluid equation. The initial electrons are provided by pre-ionization close to reality. The discharge is initiated wherever the electric field reaches a prebreakdown threshold. Here, the partial discharge starts from both ends of the particle, where the threshold is satisfied as shown in Fig.\ref{Fig.2}. The lower corona characteristics are similar to those reported by Gao \textit{et al.} \cite{15}, but the upper streamer characteristics are more complicated. Therefore, this study primarily focuses on analyzing the characteristics of upper streamer discharge.

In contrast to the single SF$_6$ streamer induced by a fixed electrode \cite{Seeger2}, the upper streamer induced by the floating particle here includes three stages, which might cause aliasing of the diagnosis signal. Stage I represents the primary streamer stage, corresponding to Figs.\ref{Fig.2}(a)--\ref{Fig.2}(c), Stage II represents the secondary streamer stage including an upward and a downward secondary streamer, corresponding to  Figs.\ref{Fig.2}(d)--\ref{Fig.2}(e), and Stage III represents the streamer extinction stage, corresponding to Fig.\ref{Fig.2}(f). 

During Stage I, since the initial floating potential of the metal particle is -35 kV, the upper streamer is similar to the negative streamer, thus there are two electron peaks positioned at the head and tail of the streamer. Another notable characteristic in the streamer channel is the presence of an electron-deficient region, which is more obvious than that in air \cite{wangzhen2022} and SF$_6$ gas mixtures \cite{reaction2,reaction3,reaction4}. This is because the positive ion region and the head electron peak together shield the channel field, making it lower than $(E/N)_{\text{cr}}$, subsequently, electrons undergo attachment and are converted into negative ions due to the strong electronegativity of SF$_6$, as shown in Fig.\ref{Fig.3}(d)--\ref{Fig.3}(f). 

Furthermore, we observe the presence of the negative net charge near the particle top tip as shown in Figs.\ref{Fig.4}(b)--\ref{Fig.4}(e), which is essentially a negative ion region and is dominated by a positive feedback mechanism. Specifically, when the tail electrons migrate into the streamer channel, they are rapidly attached due to the above low channel field and form a negative ion region outside the tail electron peak. Then the lowest field region is formed between the positive and negative ion regions, which enhances the attachment reactions there. Consequently, a positive feedback mechanism is formed, which makes it difficult for the tail electrons to escape from the lowest field region. Eventually, the tail electrons are converted to negative ions, causing the negative ion region to expand toward the top tip of the particle.

The negative ion region mentioned above plays a significant role in the recovery of the primary streamer channel field in SF$_6$ as shown in Fig.\ref{Fig.4}(h)--\ref{Fig.4}(i), which is different from that in air \cite{Nijdam_2020}. Although this field recovery phenomenon in SF$_6$ has been reported in previous studies \cite{14,sf61991}, the underlying mechanism remains unclear. Here, we report a synergistic effect of the negative ion region and head electron peak, which together determine the recovery of the channel field. On one hand, the negative ion region enhances the electric field in the streamer channel, on the other hand, as the head electron peak propagates forward, the shielding effect on the channel field away from it weakens. These two factors work together and lead to the field recovery phenomenon.

The field recovery in Stage I is the immediate cause of the upward secondary streamer in Stage II. As shown in Fig.\ref{Fig.4}(j), when the head electron peak of the primary streamer reaches the HVDC electrode, the field recovers to a level higher than $(E/N)_{\text{cr}}$. As the upward secondary ionizing wave propagates forward, the field inside the secondary channel is once again shielded below $(E/N)_{\text{cr}}$ as shown in Fig.\ref{Fig.4}(k). However, the channel field of the secondary streamer in the air has been reported to remain above its critical value \cite{Komuro_2018}. This is because the mechanisms of the secondary streamer between SF$_6$ and air are fundamentally different. Specifically, the secondary air streamer noted in the literature occurred either after breakdown or at the rising edge of the pulse voltage, which could introduce a decreasing gas density \textit{N} caused by intense heating after primary penetration \cite{secondary1984}, or operate under an increasing applied electric field \cite{Briels_2008}. However, the above two factors have little effect here, since the particle-induced partial discharge would not cause the connection of the two poles of the power supply, and the change of gas density \textit{N} can be ignored. Additionally, the effect of attachment instability which plays a key role in secondary air streamer is relatively small in SF$_6$ \cite{secondary1984}. Therefore, here the secondary streamer is essentially induced by negative ion accumulation due to the strong electronegativity of SF$_6$.

\begin{figure}[t]
\centering
\includegraphics[clip, trim=0cm 0cm 0cm 0cm]{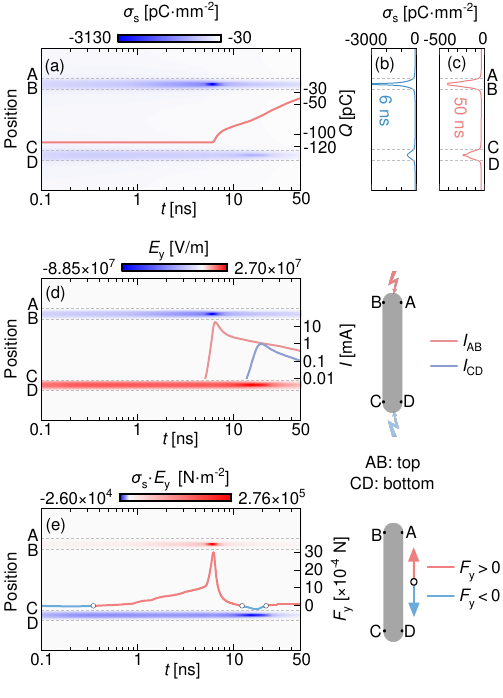}
\caption{\label{Fig.5}(a) Spatiotemporal evolution of surface charge density $\sigma_{\text{s}}$ of a metal particle and the temporal evolution of the total charge of the metal particle $Q$. The surface charge density distribution of the metal particle at (b) 6 ns and (c) 50 ns respectively. (d) The spatiotemporal evolution of vertical electric field intensity $E_{y}$ of metal particle surface and temporal evolution of conduction current on the top surface and bottom surface of the metal particle. (e) The spatiotemporal evolution of $\sigma_{\text{s}}$·$E_{y}$ of metal particle surface and temporal evolution of vertical electric field force $F_{y}$ of metal particle.}
\end{figure}
Another notable characteristic in Stage II is the increasing positive ion density on both sides of the primary residual channel after the primary streamer head contacts the HVDC electrode, as shown in Fig.\ref{Fig.3}(c). Since the time scale of Stage II is rather short for ion migration to occur, positive ions arise from ionization rather than from migration, which indicates the existence of a downward secondary streamer as shown in Figs.\ref{Fig.2}(d)--\ref{Fig.2}(e). The generation of a downward secondary streamer is mainly driven by the residual positive ions of the primary streamer near the HVDC electrode. Specifically, after the primary streamer head contacts the HVDC electrode, the electrons in the primary channel are rapidly conducted into the electrode and the residual positive ions are deposited near the electrode as shown in Fig.\ref{Fig.4}(d). However, both sides of the channel have not yet been reached and they still maintain the net negative charge characteristic of the negative streamer. Consequently, two ionization regions are formed between the residual positive ions and the net negative charges on both sides. Therefore, the downward secondary streamer occurs along both sides of the primary streamer channel rather than the residual primary channel. The upward and downward streamers inevitably interact with each other \cite{Teunissen_2017}, causing them to converge and complete the penetration of the secondary streamer stage.

During Stage III, the field in the entire channel is reduced below $(E/N)_{\text{cr}}$ as shown in Fig.\ref{Fig.4}(l), representing the end of ionization and the beginning of space charge transport, in which the interaction between space charge and metal particle constitutes the pivotal process, exerting a significant influence on the electric field force of particle. As shown in Fig.\ref{Fig.5}(d), the upper and lower discharges result in conduction currents with amplitudes of 10 mA and 1 mA at 6 ns and 20 ns respectively. The above currents mainly arise from the migration of positive ions, leading to an increase in \textit{Q} to -40 pC, as shown in Fig.\ref{Fig.5}(a). As charging progresses, $\lvert \sigma_{\text{s}} \rvert$ tends to decrease overall, as shown in Figs.\ref{Fig.5}(b)--\ref{Fig.5}(c), but it is noteworthy that during the transient process of current initiation, more $\lvert \sigma_{\text{s}} \rvert$ are generated at each end, which is related to the strong vertical electric field $E_{y}$ on the surface as shown in Fig.\ref{Fig.5}(d). Specifically, the metal particle requires the local induction of more negative charges to achieve electrostatic balance, as positive ions approach to it. This electrostatic induction phenomenon directly results in two peaks in the electric field force acting on the metal particles as shown in Fig.\ref{Fig.5}(e). However, after 20 ns, the vertical electric field force $F_{y}$ shifts from a repulsive force to a vertically upward attractive force. Notably, although $\sigma_{\text{s}}$ and \textit{Q} are both negative throughout the process, the particle can still experience an attractive $F_{y}$ toward the high voltage electrode in the negative voltage environment. This is because the direction of the vertical $F_{y}$ is not determined by \textit{Q}, but by $\sigma_{\text{s}}$·$E_{y}$ at the micro level, which is expressed as
\begin{equation}
   \label{Eq4}
F_y=\iint_{\mathrm{S}} \sigma_{\mathrm{s}} \cdot E_y \mathrm{dS}
\end{equation}
The symbols of $\sigma_{\text{s}}$·$E_{y}$  at the upper and lower ends of the particle are opposite, resulting in competition between the two ends in determining the vertical force acting on the metal particle as shown in Fig.\ref{Fig.5}(e). In general, the space charge is crucial for the force on the particle. This finding is an important supplement to the existing experimental results \cite{Chang_2023} and may offer a more precise understanding of the underlying mechanisms of the previously reported firefly phenomenon in GIL systems.

In summary, the results from this work provide an indepth understanding of the SF$_6$ partial discharge with the presence of the metal particle. The microscopic characteristics of multistage SF$_6$ discharge are notably distinct from those in air, which may lead to the aliasing of partial discharge signals. Due to the strong electronegativity of SF$_6$, a negative ion region is formed around the top tip of the metal particle, which is dominated by a positive feedback mechanism. In addition, an electric field recovery phenomenon dominated by the synergistic effect of the negative ion region and the head electron peak is reported. The subsequent upward secondary streamer is dominated by this field recovery and the downward secondary streamer is dominated by the residual space charge. Additionally, we analyze the particle charging process and the reversal of the vertical electric force, which is the dominant factor of firefly motion. The approach in this letter could have a broad application in the study of particle-induced discharge mechanisms. And the results in this letter could provide references for both detection signal and motion suppression in the GIL system. As for detection, relating Ultra High Frequency (UHF) signal can be calculated based on the current signal of discharge, with the resolution of nanosecond time, which can provide a reference for solving UHF signal aliasing. As for suppression, due to the critical role of space charge in electric force on metal particles, an "active ion source" could be a potential approach to regulate the particle movement near the electrode, so as to improve the suppression efficiency.
\vspace{1\baselineskip}
\newcommand{\sectionbackup}{\titleformat*{\section}{\Large\bfseries\centering}}
\titleformat*{\section}{\Large\bfseries\raggedright}
\section*{Acknowledgment}
The authors gratefully acknowledge the funding support from the National Natural Science Foundation of China (Contract No.~52277154). The authors thank Dr. Caomingzhe Si for fruitful discussions. The authors express sincere gratitude to the editors and reviewers for their constructive comments and suggestions, which have greatly contributed to the overall improvement of this manuscript.
\section*{AUTHOR DECLARATIONS}
\section*{Conflict of Interest} The authors have no conflicts to disclose.
\section*{DATA AVAILABILITY}
The data that support the findings of this study are available from the corresponding author upon reasonable request.
\sectionbackup
\Large
\bibliography{references}

\begin{thebibliography}{44}%
\makeatletter
\providecommand \@ifxundefined [1]{%
 \@ifx{#1\undefined}
}%
\providecommand \@ifnum [1]{%
 \ifnum #1\expandafter \@firstoftwo
 \else \expandafter \@secondoftwo
 \fi
}%
\providecommand \@ifx [1]{%
 \ifx #1\expandafter \@firstoftwo
 \else \expandafter \@secondoftwo
 \fi
}%
\providecommand \natexlab [1]{#1}%
\providecommand \enquote  [1]{``#1''}%
\providecommand \bibnamefont  [1]{#1}%
\providecommand \bibfnamefont [1]{#1}%
\providecommand \citenamefont [1]{#1}%
\providecommand \href@noop [0]{\@secondoftwo}%
\providecommand \href [0]{\begingroup \@sanitize@url \@href}%
\providecommand \@href[1]{\@@startlink{#1}\@@href}%
\providecommand \@@href[1]{\endgroup#1\@@endlink}%
\providecommand \@sanitize@url [0]{\catcode `\\12\catcode `\$12\catcode `\&12\catcode `\#12\catcode `\^12\catcode `\_12\catcode `\%12\relax}%
\providecommand \@@startlink[1]{}%
\providecommand \@@endlink[0]{}%
\providecommand \url  [0]{\begingroup\@sanitize@url \@url }%
\providecommand \@url [1]{\endgroup\@href {#1}{\urlprefix }}%
\providecommand \urlprefix  [0]{URL }%
\providecommand \Eprint [0]{\href }%
\providecommand \doibase [0]{http://dx.doi.org/}%
\providecommand \selectlanguage [0]{\@gobble}%
\providecommand \bibinfo  [0]{\@secondoftwo}%
\providecommand \bibfield  [0]{\@secondoftwo}%
\providecommand \translation [1]{[#1]}%
\providecommand \BibitemOpen [0]{}%
\providecommand \bibitemStop [0]{}%
\providecommand \bibitemNoStop [0]{.\EOS\space}%
\providecommand \EOS [0]{\spacefactor3000\relax}%
\providecommand \BibitemShut  [1]{\csname bibitem#1\endcsname}%
\let\auto@bib@innerbib\@empty
\bibitem [{\citenamefont {Li}\ \emph {et~al.}(2022)\citenamefont {Li}, \citenamefont {Zhang}, \citenamefont {Lv}, \citenamefont {Liang}, \citenamefont {Liang}, \citenamefont {Fan}, \citenamefont {Riechert}, \citenamefont {Li}, \citenamefont {Liu}, \citenamefont {Xue}, \citenamefont {Pan}, \citenamefont {Chen}, \citenamefont {Zhang}, \citenamefont {Wang}, \citenamefont {Lu}, \citenamefont {Liang}, \citenamefont {Pan}, \citenamefont {Zhuang}, \citenamefont {Mazzanti}, \citenamefont {Fabiani}, \citenamefont {Liu}, \citenamefont {Cao}, \citenamefont {Zhong}, \citenamefont {Deng}, \citenamefont {Nan}, \citenamefont {Tang},\ and\ \citenamefont {He}}]{10007898}%
  \BibitemOpen
  \bibfield  {author} {\bibinfo {author} {\bibfnamefont {C.}~\bibnamefont {Li}}, \bibinfo {author} {\bibfnamefont {C.}~\bibnamefont {Zhang}}, \bibinfo {author} {\bibfnamefont {J.}~\bibnamefont {Lv}}, \bibinfo {author} {\bibfnamefont {F.}~\bibnamefont {Liang}}, \bibinfo {author} {\bibfnamefont {Z.}~\bibnamefont {Liang}}, \bibinfo {author} {\bibfnamefont {X.}~\bibnamefont {Fan}}, \bibinfo {author} {\bibfnamefont {U.}~\bibnamefont {Riechert}}, \bibinfo {author} {\bibfnamefont {Z.}~\bibnamefont {Li}}, \bibinfo {author} {\bibfnamefont {P.}~\bibnamefont {Liu}}, \bibinfo {author} {\bibfnamefont {J.}~\bibnamefont {Xue}}, \bibinfo {author} {\bibfnamefont {C.}~\bibnamefont {Pan}}, \bibinfo {author} {\bibfnamefont {G.}~\bibnamefont {Chen}}, \bibinfo {author} {\bibfnamefont {L.}~\bibnamefont {Zhang}}, \bibinfo {author} {\bibfnamefont {Z.}~\bibnamefont {Wang}}, \bibinfo {author} {\bibfnamefont {W.}~\bibnamefont {Lu}}, \bibinfo {author} {\bibfnamefont {H.}~\bibnamefont {Liang}}, \bibinfo {author} {\bibfnamefont
  {Z.}~\bibnamefont {Pan}}, \bibinfo {author} {\bibfnamefont {W.}~\bibnamefont {Zhuang}}, \bibinfo {author} {\bibfnamefont {G.}~\bibnamefont {Mazzanti}}, \bibinfo {author} {\bibfnamefont {D.}~\bibnamefont {Fabiani}}, \bibinfo {author} {\bibfnamefont {B.}~\bibnamefont {Liu}}, \bibinfo {author} {\bibfnamefont {S.}~\bibnamefont {Cao}}, \bibinfo {author} {\bibfnamefont {J.}~\bibnamefont {Zhong}}, \bibinfo {author} {\bibfnamefont {Y.}~\bibnamefont {Deng}}, \bibinfo {author} {\bibfnamefont {Z.}~\bibnamefont {Nan}}, \bibinfo {author} {\bibfnamefont {J.}~\bibnamefont {Tang}}, \ and\ \bibinfo {author} {\bibfnamefont {J.}~\bibnamefont {He}},\ }\href {\doibase 10.23919/IEN.2022.0050} {\bibfield  {journal} {\bibinfo  {journal} {iEnergy}\ }\textbf {\bibinfo {volume} {1}},\ \bibinfo {pages} {400} (\bibinfo {year} {2022})}\BibitemShut {NoStop}%
\bibitem [{\citenamefont {Zhuang}\ \emph {et~al.}(2023{\natexlab{a}})\citenamefont {Zhuang}, \citenamefont {Liang}, \citenamefont {Liang}, \citenamefont {Zhang}, \citenamefont {Li}, \citenamefont {Zhang},\ and\ \citenamefont {He}}]{trap}%
  \BibitemOpen
  \bibfield  {author} {\bibinfo {author} {\bibfnamefont {W.}~\bibnamefont {Zhuang}}, \bibinfo {author} {\bibfnamefont {F.}~\bibnamefont {Liang}}, \bibinfo {author} {\bibfnamefont {Z.}~\bibnamefont {Liang}}, \bibinfo {author} {\bibfnamefont {C.}~\bibnamefont {Zhang}}, \bibinfo {author} {\bibfnamefont {C.}~\bibnamefont {Li}}, \bibinfo {author} {\bibfnamefont {B.}~\bibnamefont {Zhang}}, \ and\ \bibinfo {author} {\bibfnamefont {J.}~\bibnamefont {He}},\ }\href {\doibase 10.1109/TDEI.2023.3343672} {\bibfield  {journal} {\bibinfo  {journal} {IEEE Trans. Dielectr. Electr. Insul.}\ ,\ \bibinfo {pages} {1}} (\bibinfo {year} {2023}{\natexlab{a}})}\BibitemShut {NoStop}%
\bibitem [{\citenamefont {Ma}\ \emph {et~al.}(2018)\citenamefont {Ma}, \citenamefont {Zhang}, \citenamefont {Wu}, \citenamefont {Guo}, \citenamefont {Wen}, \citenamefont {Wang},\ and\ \citenamefont {Gao}}]{exp3}%
  \BibitemOpen
  \bibfield  {author} {\bibinfo {author} {\bibfnamefont {J.}~\bibnamefont {Ma}}, \bibinfo {author} {\bibfnamefont {Q.}~\bibnamefont {Zhang}}, \bibinfo {author} {\bibfnamefont {Z.}~\bibnamefont {Wu}}, \bibinfo {author} {\bibfnamefont {C.}~\bibnamefont {Guo}}, \bibinfo {author} {\bibfnamefont {T.}~\bibnamefont {Wen}}, \bibinfo {author} {\bibfnamefont {G.}~\bibnamefont {Wang}}, \ and\ \bibinfo {author} {\bibfnamefont {C.}~\bibnamefont {Gao}},\ }\href {\doibase 10.1109/TDEI.2018.007053} {\bibfield  {journal} {\bibinfo  {journal} {IEEE Trans. Dielectr. Electr. Insul.}\ }\textbf {\bibinfo {volume} {25}},\ \bibinfo {pages} {1439} (\bibinfo {year} {2018})}\BibitemShut {NoStop}%
\bibitem [{\citenamefont {Sakai}\ \emph {et~al.}(1999)\citenamefont {Sakai}, \citenamefont {Tsuru}, \citenamefont {Abella},\ and\ \citenamefont {Hara}}]{exp4}%
  \BibitemOpen
  \bibfield  {author} {\bibinfo {author} {\bibfnamefont {K.}~\bibnamefont {Sakai}}, \bibinfo {author} {\bibfnamefont {S.}~\bibnamefont {Tsuru}}, \bibinfo {author} {\bibfnamefont {D.}~\bibnamefont {Abella}}, \ and\ \bibinfo {author} {\bibfnamefont {M.}~\bibnamefont {Hara}},\ }\href {\doibase 10.1109/94.752020} {\bibfield  {journal} {\bibinfo  {journal} {IEEE Trans. Dielectr. Electr. Insul.}\ }\textbf {\bibinfo {volume} {6}},\ \bibinfo {pages} {122} (\bibinfo {year} {1999})}\BibitemShut {NoStop}%
\bibitem [{\citenamefont {Beckers}\ \emph {et~al.}(2023)\citenamefont {Beckers}, \citenamefont {Berndt}, \citenamefont {Block}, \citenamefont {Bonitz}, \citenamefont {Bruggeman}, \citenamefont {Couëdel}, \citenamefont {Delzanno}, \citenamefont {Feng}, \citenamefont {Gopalakrishnan}, \citenamefont {Greiner}, \citenamefont {Hartmann}, \citenamefont {Horányi}, \citenamefont {Kersten}, \citenamefont {Knapek}, \citenamefont {Konopka}, \citenamefont {Kortshagen}, \citenamefont {Kostadinova}, \citenamefont {Kovačević}, \citenamefont {Krasheninnikov}, \citenamefont {Mann}, \citenamefont {Mariotti}, \citenamefont {Matthews}, \citenamefont {Melzer}, \citenamefont {Mikikian}, \citenamefont {Nosenko}, \citenamefont {Pustylnik}, \citenamefont {Ratynskaia}, \citenamefont {Sankaran}, \citenamefont {Schneider}, \citenamefont {Thimsen}, \citenamefont {Thomas}, \citenamefont {Thomas}, \citenamefont {Tolias},\ and\ \citenamefont {van~de Kerkhof}}]{26}%
  \BibitemOpen
  \bibfield  {author} {\bibinfo {author} {\bibfnamefont {J.}~\bibnamefont {Beckers}}, \bibinfo {author} {\bibfnamefont {J.}~\bibnamefont {Berndt}}, \bibinfo {author} {\bibfnamefont {D.}~\bibnamefont {Block}}, \bibinfo {author} {\bibfnamefont {M.}~\bibnamefont {Bonitz}}, \bibinfo {author} {\bibfnamefont {P.~J.}\ \bibnamefont {Bruggeman}}, \bibinfo {author} {\bibfnamefont {L.}~\bibnamefont {Couëdel}}, \bibinfo {author} {\bibfnamefont {G.~L.}\ \bibnamefont {Delzanno}}, \bibinfo {author} {\bibfnamefont {Y.}~\bibnamefont {Feng}}, \bibinfo {author} {\bibfnamefont {R.}~\bibnamefont {Gopalakrishnan}}, \bibinfo {author} {\bibfnamefont {F.}~\bibnamefont {Greiner}}, \bibinfo {author} {\bibfnamefont {P.}~\bibnamefont {Hartmann}}, \bibinfo {author} {\bibfnamefont {M.}~\bibnamefont {Horányi}}, \bibinfo {author} {\bibfnamefont {H.}~\bibnamefont {Kersten}}, \bibinfo {author} {\bibfnamefont {C.~A.}\ \bibnamefont {Knapek}}, \bibinfo {author} {\bibfnamefont {U.}~\bibnamefont {Konopka}}, \bibinfo {author} {\bibfnamefont
  {U.}~\bibnamefont {Kortshagen}}, \bibinfo {author} {\bibfnamefont {E.~G.}\ \bibnamefont {Kostadinova}}, \bibinfo {author} {\bibfnamefont {E.}~\bibnamefont {Kovačević}}, \bibinfo {author} {\bibfnamefont {S.~I.}\ \bibnamefont {Krasheninnikov}}, \bibinfo {author} {\bibfnamefont {I.}~\bibnamefont {Mann}}, \bibinfo {author} {\bibfnamefont {D.}~\bibnamefont {Mariotti}}, \bibinfo {author} {\bibfnamefont {L.~S.}\ \bibnamefont {Matthews}}, \bibinfo {author} {\bibfnamefont {A.}~\bibnamefont {Melzer}}, \bibinfo {author} {\bibfnamefont {M.}~\bibnamefont {Mikikian}}, \bibinfo {author} {\bibfnamefont {V.}~\bibnamefont {Nosenko}}, \bibinfo {author} {\bibfnamefont {M.~Y.}\ \bibnamefont {Pustylnik}}, \bibinfo {author} {\bibfnamefont {S.}~\bibnamefont {Ratynskaia}}, \bibinfo {author} {\bibfnamefont {R.~M.}\ \bibnamefont {Sankaran}}, \bibinfo {author} {\bibfnamefont {V.}~\bibnamefont {Schneider}}, \bibinfo {author} {\bibfnamefont {E.~J.}\ \bibnamefont {Thimsen}}, \bibinfo {author} {\bibfnamefont {E.}~\bibnamefont {Thomas}},
  \bibinfo {author} {\bibfnamefont {H.~M.}\ \bibnamefont {Thomas}}, \bibinfo {author} {\bibfnamefont {P.}~\bibnamefont {Tolias}}, \ and\ \bibinfo {author} {\bibfnamefont {M.}~\bibnamefont {van~de Kerkhof}},\ }\href {\doibase 10.1063/5.0168088} {\bibfield  {journal} {\bibinfo  {journal} {Phys. Plasmas}\ }\textbf {\bibinfo {volume} {30}},\ \bibinfo {pages} {120601} (\bibinfo {year} {2023})}\BibitemShut {NoStop}%
\bibitem [{\citenamefont {Chen}\ \emph {et~al.}(2022)\citenamefont {Chen}, \citenamefont {Verboncoeur},\ and\ \citenamefont {Fu}}]{10.1063/5.0104205}%
  \BibitemOpen
  \bibfield  {author} {\bibinfo {author} {\bibfnamefont {J.}~\bibnamefont {Chen}}, \bibinfo {author} {\bibfnamefont {J.~P.}\ \bibnamefont {Verboncoeur}}, \ and\ \bibinfo {author} {\bibfnamefont {Y.}~\bibnamefont {Fu}},\ }\href {\doibase 10.1063/5.0104205} {\bibfield  {journal} {\bibinfo  {journal} {Appl. Phys. Lett.}\ }\textbf {\bibinfo {volume} {121}},\ \bibinfo {pages} {074102} (\bibinfo {year} {2022})}\BibitemShut {NoStop}%
\bibitem [{\citenamefont {Sun}\ \emph {et~al.}(2021)\citenamefont {Sun}, \citenamefont {hong Zhou}, \citenamefont {Yang}, \citenamefont {Dong}, \citenamefont {tian Zhang}, \citenamefont {meng Song},\ and\ \citenamefont {Wu}}]{Sun_2021}%
  \BibitemOpen
  \bibfield  {author} {\bibinfo {author} {\bibfnamefont {Q.}~\bibnamefont {Sun}}, \bibinfo {author} {\bibfnamefont {Q.}~\bibnamefont {hong Zhou}}, \bibinfo {author} {\bibfnamefont {W.}~\bibnamefont {Yang}}, \bibinfo {author} {\bibfnamefont {Y.}~\bibnamefont {Dong}}, \bibinfo {author} {\bibfnamefont {H.}~\bibnamefont {tian Zhang}}, \bibinfo {author} {\bibfnamefont {M.}~\bibnamefont {meng Song}}, \ and\ \bibinfo {author} {\bibfnamefont {Y.}~\bibnamefont {Wu}},\ }\href {\doibase 10.1088/1361-6595/abec26} {\bibfield  {journal} {\bibinfo  {journal} {Plasma Sources Sci. Technol.}\ }\textbf {\bibinfo {volume} {30}},\ \bibinfo {pages} {045001} (\bibinfo {year} {2021})}\BibitemShut {NoStop}%
\bibitem [{\citenamefont {Zhong}\ \emph {et~al.}(2022)\citenamefont {Zhong}, \citenamefont {Xu}, \citenamefont {Shi}, \citenamefont {Jin},\ and\ \citenamefont {Chen}}]{https://doi.org/10.1002/ctpp.202100133}%
  \BibitemOpen
  \bibfield  {author} {\bibinfo {author} {\bibfnamefont {W.}~\bibnamefont {Zhong}}, \bibinfo {author} {\bibfnamefont {A.}~\bibnamefont {Xu}}, \bibinfo {author} {\bibfnamefont {Y.}~\bibnamefont {Shi}}, \bibinfo {author} {\bibfnamefont {D.}~\bibnamefont {Jin}}, \ and\ \bibinfo {author} {\bibfnamefont {L.}~\bibnamefont {Chen}},\ }\href {\doibase https://doi.org/10.1002/ctpp.202100133} {\bibfield  {journal} {\bibinfo  {journal} {Contrib. Plasma Phys.}\ }\textbf {\bibinfo {volume} {62}},\ \bibinfo {pages} {e202100133} (\bibinfo {year} {2022})}\BibitemShut {NoStop}%
\bibitem [{\citenamefont {Gallimberti}\ and\ \citenamefont {Wiegart}(1986)}]{SF6111}%
  \BibitemOpen
  \bibfield  {author} {\bibinfo {author} {\bibfnamefont {I.}~\bibnamefont {Gallimberti}}\ and\ \bibinfo {author} {\bibfnamefont {N.}~\bibnamefont {Wiegart}},\ }\href {\doibase 10.1088/0022-3727/19/12/016} {\bibfield  {journal} {\bibinfo  {journal} {J. Phys. D: Appl. Phys.}\ }\textbf {\bibinfo {volume} {19}},\ \bibinfo {pages} {2363} (\bibinfo {year} {1986})}\BibitemShut {NoStop}%
\bibitem [{\citenamefont {Morrow}(1987)}]{sf6333}%
  \BibitemOpen
  \bibfield  {author} {\bibinfo {author} {\bibfnamefont {R.}~\bibnamefont {Morrow}},\ }\href {\doibase 10.1103/PhysRevA.35.1778} {\bibfield  {journal} {\bibinfo  {journal} {Phys. Rev. A}\ }\textbf {\bibinfo {volume} {35}},\ \bibinfo {pages} {1778} (\bibinfo {year} {1987})}\BibitemShut {NoStop}%
\bibitem [{\citenamefont {Dhali}\ and\ \citenamefont {Pal}(1988)}]{14}%
  \BibitemOpen
  \bibfield  {author} {\bibinfo {author} {\bibfnamefont {S.~K.}\ \bibnamefont {Dhali}}\ and\ \bibinfo {author} {\bibfnamefont {A.~K.}\ \bibnamefont {Pal}},\ }\href {\doibase 10.1063/1.339963} {\bibfield  {journal} {\bibinfo  {journal} {J. Appl. Phys.}\ }\textbf {\bibinfo {volume} {63}},\ \bibinfo {pages} {1355} (\bibinfo {year} {1988})}\BibitemShut {NoStop}%
\bibitem [{\citenamefont {Morrow}(1991)}]{sf61991}%
  \BibitemOpen
  \bibfield  {author} {\bibinfo {author} {\bibfnamefont {R.}~\bibnamefont {Morrow}},\ }\href {\doibase 10.1109/27.106801} {\bibfield  {journal} {\bibinfo  {journal} {IEEE Trans. Plasma Sci.}\ }\textbf {\bibinfo {volume} {19}},\ \bibinfo {pages} {86} (\bibinfo {year} {1991})}\BibitemShut {NoStop}%
\bibitem [{\citenamefont {Espel}\ \emph {et~al.}(2002)\citenamefont {Espel}, \citenamefont {Paillol}, \citenamefont {Reess}, \citenamefont {Gibert},\ and\ \citenamefont {Domens}}]{SF62002}%
  \BibitemOpen
  \bibfield  {author} {\bibinfo {author} {\bibfnamefont {P.}~\bibnamefont {Espel}}, \bibinfo {author} {\bibfnamefont {J.}~\bibnamefont {Paillol}}, \bibinfo {author} {\bibfnamefont {T.}~\bibnamefont {Reess}}, \bibinfo {author} {\bibfnamefont {A.}~\bibnamefont {Gibert}}, \ and\ \bibinfo {author} {\bibfnamefont {P.}~\bibnamefont {Domens}},\ }\href {\doibase 10.1088/0022-3727/35/4/307} {\bibfield  {journal} {\bibinfo  {journal} {J. Phys. D: Appl. Phys.}\ }\textbf {\bibinfo {volume} {35}},\ \bibinfo {pages} {318} (\bibinfo {year} {2002})}\BibitemShut {NoStop}%
\bibitem [{\citenamefont {Gao}\ \emph {et~al.}(2018)\citenamefont {Gao}, \citenamefont {Niu}, \citenamefont {Adamiak}, \citenamefont {Yang}, \citenamefont {Rong},\ and\ \citenamefont {Wang}}]{15}%
  \BibitemOpen
  \bibfield  {author} {\bibinfo {author} {\bibfnamefont {Q.}~\bibnamefont {Gao}}, \bibinfo {author} {\bibfnamefont {C.}~\bibnamefont {Niu}}, \bibinfo {author} {\bibfnamefont {K.}~\bibnamefont {Adamiak}}, \bibinfo {author} {\bibfnamefont {A.}~\bibnamefont {Yang}}, \bibinfo {author} {\bibfnamefont {M.}~\bibnamefont {Rong}}, \ and\ \bibinfo {author} {\bibfnamefont {X.}~\bibnamefont {Wang}},\ }\href {\doibase 10.1088/1361-6595/aae706} {\bibfield  {journal} {\bibinfo  {journal} {Plasma Sources Sci. Technol.}\ }\textbf {\bibinfo {volume} {27}},\ \bibinfo {pages} {115001} (\bibinfo {year} {2018})}\BibitemShut {NoStop}%
\bibitem [{\citenamefont {Yang}\ \emph {et~al.}(2022)\citenamefont {Yang}, \citenamefont {Zhu}, \citenamefont {Gao}, \citenamefont {Du}, \citenamefont {Zeng},\ and\ \citenamefont {Zhang}}]{19}%
  \BibitemOpen
  \bibfield  {author} {\bibinfo {author} {\bibfnamefont {D.}~\bibnamefont {Yang}}, \bibinfo {author} {\bibfnamefont {L.}~\bibnamefont {Zhu}}, \bibinfo {author} {\bibfnamefont {Y.}~\bibnamefont {Gao}}, \bibinfo {author} {\bibfnamefont {H.}~\bibnamefont {Du}}, \bibinfo {author} {\bibfnamefont {F.}~\bibnamefont {Zeng}}, \ and\ \bibinfo {author} {\bibfnamefont {G.}~\bibnamefont {Zhang}},\ }\href {\doibase 10.1063/5.0086498} {\bibfield  {journal} {\bibinfo  {journal} {AIP Adv.}\ }\textbf {\bibinfo {volume} {12}},\ \bibinfo {pages} {045226} (\bibinfo {year} {2022})}\BibitemShut {NoStop}%
\bibitem [{\citenamefont {Zhuang}\ \emph {et~al.}(2023{\natexlab{b}})\citenamefont {Zhuang}, \citenamefont {Liang}, \citenamefont {Liang}, \citenamefont {Fan}, \citenamefont {Luo}, \citenamefont {Hu}, \citenamefont {Li}, \citenamefont {Zhang},\ and\ \citenamefont {He}}]{Zhuang_2023}%
  \BibitemOpen
  \bibfield  {author} {\bibinfo {author} {\bibfnamefont {W.}~\bibnamefont {Zhuang}}, \bibinfo {author} {\bibfnamefont {Z.}~\bibnamefont {Liang}}, \bibinfo {author} {\bibfnamefont {F.}~\bibnamefont {Liang}}, \bibinfo {author} {\bibfnamefont {X.}~\bibnamefont {Fan}}, \bibinfo {author} {\bibfnamefont {H.}~\bibnamefont {Luo}}, \bibinfo {author} {\bibfnamefont {J.}~\bibnamefont {Hu}}, \bibinfo {author} {\bibfnamefont {C.}~\bibnamefont {Li}}, \bibinfo {author} {\bibfnamefont {B.}~\bibnamefont {Zhang}}, \ and\ \bibinfo {author} {\bibfnamefont {J.}~\bibnamefont {He}},\ }\href {\doibase 10.1088/1361-6463/accfa8} {\bibfield  {journal} {\bibinfo  {journal} {J. Phys. D: Appl. Phys.}\ }\textbf {\bibinfo {volume} {56}},\ \bibinfo {pages} {325501} (\bibinfo {year} {2023}{\natexlab{b}})}\BibitemShut {NoStop}%
\bibitem [{\citenamefont {Diessner}\ and\ \citenamefont {Trump}(1970)}]{4074282}%
  \BibitemOpen
  \bibfield  {author} {\bibinfo {author} {\bibfnamefont {A.}~\bibnamefont {Diessner}}\ and\ \bibinfo {author} {\bibfnamefont {J.~G.}\ \bibnamefont {Trump}},\ }\href {\doibase 10.1109/TPAS.1970.292781} {\bibfield  {journal} {\bibinfo  {journal} {IEEE Trans. Power App. Syst.}\ }\textbf {\bibinfo {volume} {PAS-89}},\ \bibinfo {pages} {1970} (\bibinfo {year} {1970})}\BibitemShut {NoStop}%
\bibitem [{\citenamefont {Chang}\ \emph {et~al.}(2023)\citenamefont {Chang}, \citenamefont {Geng}, \citenamefont {Hu}, \citenamefont {Li}, \citenamefont {Li},\ and\ \citenamefont {Wang}}]{Chang_2023}%
  \BibitemOpen
  \bibfield  {author} {\bibinfo {author} {\bibfnamefont {Y.}~\bibnamefont {Chang}}, \bibinfo {author} {\bibfnamefont {Q.}~\bibnamefont {Geng}}, \bibinfo {author} {\bibfnamefont {Z.}~\bibnamefont {Hu}}, \bibinfo {author} {\bibfnamefont {Z.}~\bibnamefont {Li}}, \bibinfo {author} {\bibfnamefont {Q.}~\bibnamefont {Li}}, \ and\ \bibinfo {author} {\bibfnamefont {J.}~\bibnamefont {Wang}},\ }\href {\doibase 10.1088/1402-4896/acdeb2} {\bibfield  {journal} {\bibinfo  {journal} {Phys. Scr.}\ }\textbf {\bibinfo {volume} {98}},\ \bibinfo {pages} {075939} (\bibinfo {year} {2023})}\BibitemShut {NoStop}%
\bibitem [{\citenamefont {Sima}\ \emph {et~al.}(2012)\citenamefont {Sima}, \citenamefont {Peng}, \citenamefont {Yang}, \citenamefont {Yuan},\ and\ \citenamefont {Shi}}]{equation}%
  \BibitemOpen
  \bibfield  {author} {\bibinfo {author} {\bibfnamefont {W.}~\bibnamefont {Sima}}, \bibinfo {author} {\bibfnamefont {Q.}~\bibnamefont {Peng}}, \bibinfo {author} {\bibfnamefont {Q.}~\bibnamefont {Yang}}, \bibinfo {author} {\bibfnamefont {T.}~\bibnamefont {Yuan}}, \ and\ \bibinfo {author} {\bibfnamefont {J.}~\bibnamefont {Shi}},\ }\href {\doibase 10.1109/TDEI.2012.6180261} {\bibfield  {journal} {\bibinfo  {journal} {IEEE Trans. Dielectr. Electr. Insul.}\ }\textbf {\bibinfo {volume} {19}},\ \bibinfo {pages} {660} (\bibinfo {year} {2012})}\BibitemShut {NoStop}%
\bibitem [{\citenamefont {Zhang}\ \emph {et~al.}(2023)\citenamefont {Zhang}, \citenamefont {Song}, \citenamefont {Luo}, \citenamefont {Sheng},\ and\ \citenamefont {Jiang}}]{16}%
  \BibitemOpen
  \bibfield  {author} {\bibinfo {author} {\bibfnamefont {Z.}~\bibnamefont {Zhang}}, \bibinfo {author} {\bibfnamefont {H.}~\bibnamefont {Song}}, \bibinfo {author} {\bibfnamefont {L.}~\bibnamefont {Luo}}, \bibinfo {author} {\bibfnamefont {G.}~\bibnamefont {Sheng}}, \ and\ \bibinfo {author} {\bibfnamefont {X.}~\bibnamefont {Jiang}},\ }\href {\doibase 10.1109/TDEI.2023.3264960} {\bibfield  {journal} {\bibinfo  {journal} {IEEE Trans. Dielectr. Electr. Insul.}\ }\textbf {\bibinfo {volume} {30}},\ \bibinfo {pages} {1769} (\bibinfo {year} {2023})}\BibitemShut {NoStop}%
\bibitem [{\citenamefont {Luo}\ \emph {et~al.}(2020)\citenamefont {Luo}, \citenamefont {He}, \citenamefont {Cheng}, \citenamefont {Xia}, \citenamefont {Du}, \citenamefont {Bian},\ and\ \citenamefont {Chen}}]{reaction2}%
  \BibitemOpen
  \bibfield  {author} {\bibinfo {author} {\bibfnamefont {B.}~\bibnamefont {Luo}}, \bibinfo {author} {\bibfnamefont {H.}~\bibnamefont {He}}, \bibinfo {author} {\bibfnamefont {C.}~\bibnamefont {Cheng}}, \bibinfo {author} {\bibfnamefont {S.}~\bibnamefont {Xia}}, \bibinfo {author} {\bibfnamefont {W.}~\bibnamefont {Du}}, \bibinfo {author} {\bibfnamefont {K.}~\bibnamefont {Bian}}, \ and\ \bibinfo {author} {\bibfnamefont {W.}~\bibnamefont {Chen}},\ }\href {\doibase 10.1109/TDEI.2019.008551} {\bibfield  {journal} {\bibinfo  {journal} {IEEE Trans. Dielectr. Electr. Insul.}\ }\textbf {\bibinfo {volume} {27}},\ \bibinfo {pages} {782} (\bibinfo {year} {2020})}\BibitemShut {NoStop}%
\bibitem [{\citenamefont {Ou}\ \emph {et~al.}(2020)\citenamefont {Ou}, \citenamefont {Wang}, \citenamefont {Liu},\ and\ \citenamefont {Lin}}]{reaction3}%
  \BibitemOpen
  \bibfield  {author} {\bibinfo {author} {\bibfnamefont {X.}~\bibnamefont {Ou}}, \bibinfo {author} {\bibfnamefont {L.}~\bibnamefont {Wang}}, \bibinfo {author} {\bibfnamefont {J.}~\bibnamefont {Liu}}, \ and\ \bibinfo {author} {\bibfnamefont {X.}~\bibnamefont {Lin}},\ }\href {\doibase 10.1063/5.0006140} {\bibfield  {journal} {\bibinfo  {journal} {Phys. Plasmas}\ }\textbf {\bibinfo {volume} {27}},\ \bibinfo {pages} {073504} (\bibinfo {year} {2020})}\BibitemShut {NoStop}%
\bibitem [{\citenamefont {Hagelaar}\ and\ \citenamefont {Pitchford}(2005)}]{bosig+}%
  \BibitemOpen
  \bibfield  {author} {\bibinfo {author} {\bibfnamefont {G.~J.~M.}\ \bibnamefont {Hagelaar}}\ and\ \bibinfo {author} {\bibfnamefont {L.~C.}\ \bibnamefont {Pitchford}},\ }\href {\doibase 10.1088/0963-0252/14/4/011} {\bibfield  {journal} {\bibinfo  {journal} {Plasma Sources Sci. Technol.}\ }\textbf {\bibinfo {volume} {14}},\ \bibinfo {pages} {722} (\bibinfo {year} {2005})}\BibitemShut {NoStop}%
\bibitem [{\citenamefont {Christophorou}\ and\ \citenamefont {Olthoff}(2000)}]{EN3601}%
  \BibitemOpen
  \bibfield  {author} {\bibinfo {author} {\bibfnamefont {L.~G.}\ \bibnamefont {Christophorou}}\ and\ \bibinfo {author} {\bibfnamefont {J.~K.}\ \bibnamefont {Olthoff}},\ }\href {\doibase 10.1063/1.1288407} {\bibfield  {journal} {\bibinfo  {journal} {J. Phys. Chem. Ref. Data}\ }\textbf {\bibinfo {volume} {29}},\ \bibinfo {pages} {267} (\bibinfo {year} {2000})}\BibitemShut {NoStop}%
\bibitem [{\citenamefont {Morrow}(1986)}]{EN3602}%
  \BibitemOpen
  \bibfield  {author} {\bibinfo {author} {\bibfnamefont {R.}~\bibnamefont {Morrow}},\ }\href {\doibase 10.1109/TPS.1986.4316534} {\bibfield  {journal} {\bibinfo  {journal} {IEEE Trans. Plasma Sci.}\ }\textbf {\bibinfo {volume} {14}},\ \bibinfo {pages} {234} (\bibinfo {year} {1986})}\BibitemShut {NoStop}%
\bibitem [{\citenamefont {Wang}\ \emph {et~al.}(2016)\citenamefont {Wang}, \citenamefont {Li}, \citenamefont {Li}, \citenamefont {Chen}, \citenamefont {Liu},\ and\ \citenamefont {Li}}]{expscaled2}%
  \BibitemOpen
  \bibfield  {author} {\bibinfo {author} {\bibfnamefont {J.}~\bibnamefont {Wang}}, \bibinfo {author} {\bibfnamefont {Q.}~\bibnamefont {Li}}, \bibinfo {author} {\bibfnamefont {B.}~\bibnamefont {Li}}, \bibinfo {author} {\bibfnamefont {C.}~\bibnamefont {Chen}}, \bibinfo {author} {\bibfnamefont {S.}~\bibnamefont {Liu}}, \ and\ \bibinfo {author} {\bibfnamefont {C.}~\bibnamefont {Li}},\ }\href {\doibase 10.1109/TDEI.2016.7556466} {\bibfield  {journal} {\bibinfo  {journal} {IEEE Trans. Dielectr. Electr. Insul.}\ }\textbf {\bibinfo {volume} {23}},\ \bibinfo {pages} {1951} (\bibinfo {year} {2016})}\BibitemShut {NoStop}%
\bibitem [{\citenamefont {Wu}\ \emph {et~al.}(2019{\natexlab{a}})\citenamefont {Wu}, \citenamefont {Zhang}, \citenamefont {Song},\ and\ \citenamefont {Li}}]{expscaled3}%
  \BibitemOpen
  \bibfield  {author} {\bibinfo {author} {\bibfnamefont {Z.}~\bibnamefont {Wu}}, \bibinfo {author} {\bibfnamefont {Q.}~\bibnamefont {Zhang}}, \bibinfo {author} {\bibfnamefont {J.}~\bibnamefont {Song}}, \ and\ \bibinfo {author} {\bibfnamefont {X.}~\bibnamefont {Li}},\ }\href {\doibase 10.1109/TPWRD.2018.2888597} {\bibfield  {journal} {\bibinfo  {journal} {IEEE Trans. Power Deliv.}\ }\textbf {\bibinfo {volume} {34}},\ \bibinfo {pages} {1317} (\bibinfo {year} {2019}{\natexlab{a}})}\BibitemShut {NoStop}%
\bibitem [{\citenamefont {Sun}\ \emph {et~al.}(2018)\citenamefont {Sun}, \citenamefont {Chen}, \citenamefont {Bian}, \citenamefont {Li}, \citenamefont {Yan},\ and\ \citenamefont {Xu}}]{particularregion}%
  \BibitemOpen
  \bibfield  {author} {\bibinfo {author} {\bibfnamefont {J.}~\bibnamefont {Sun}}, \bibinfo {author} {\bibfnamefont {W.}~\bibnamefont {Chen}}, \bibinfo {author} {\bibfnamefont {K.}~\bibnamefont {Bian}}, \bibinfo {author} {\bibfnamefont {Z.}~\bibnamefont {Li}}, \bibinfo {author} {\bibfnamefont {X.}~\bibnamefont {Yan}}, \ and\ \bibinfo {author} {\bibfnamefont {Y.}~\bibnamefont {Xu}},\ }\href {\doibase 10.1109/TDEI.2018.006879} {\bibfield  {journal} {\bibinfo  {journal} {IEEE Trans. Dielectr. Electr. Insul.}\ }\textbf {\bibinfo {volume} {25}},\ \bibinfo {pages} {1047} (\bibinfo {year} {2018})}\BibitemShut {NoStop}%
\bibitem [{\citenamefont {Hillhouse}\ and\ \citenamefont {Peterson}(1973)}]{radio}%
  \BibitemOpen
  \bibfield  {author} {\bibinfo {author} {\bibfnamefont {D.~L.}\ \bibnamefont {Hillhouse}}\ and\ \bibinfo {author} {\bibfnamefont {A.~E.}\ \bibnamefont {Peterson}},\ }\href {\doibase 10.1109/TIM.1973.4314197} {\bibfield  {journal} {\bibinfo  {journal} {IEEE Trans. Instrum. Meas.}\ }\textbf {\bibinfo {volume} {22}},\ \bibinfo {pages} {408} (\bibinfo {year} {1973})}\BibitemShut {NoStop}%
\bibitem [{\citenamefont {Wenger}\ \emph {et~al.}(2019)\citenamefont {Wenger}, \citenamefont {Beltle}, \citenamefont {Tenbohlen}, \citenamefont {Riechert},\ and\ \citenamefont {Behrmann}}]{8684218}%
  \BibitemOpen
  \bibfield  {author} {\bibinfo {author} {\bibfnamefont {P.}~\bibnamefont {Wenger}}, \bibinfo {author} {\bibfnamefont {M.}~\bibnamefont {Beltle}}, \bibinfo {author} {\bibfnamefont {S.}~\bibnamefont {Tenbohlen}}, \bibinfo {author} {\bibfnamefont {U.}~\bibnamefont {Riechert}}, \ and\ \bibinfo {author} {\bibfnamefont {G.}~\bibnamefont {Behrmann}},\ }\href {\doibase 10.1109/TPWRD.2019.2909830} {\bibfield  {journal} {\bibinfo  {journal} {IEEE Trans. Power Deliv.}\ }\textbf {\bibinfo {volume} {34}},\ \bibinfo {pages} {1540} (\bibinfo {year} {2019})}\BibitemShut {NoStop}%
\bibitem [{\citenamefont {Pedersen}(1970)}]{pd}%
  \BibitemOpen
  \bibfield  {author} {\bibinfo {author} {\bibfnamefont {A.}~\bibnamefont {Pedersen}},\ }\href {\doibase 10.1109/TPAS.1970.292789} {\bibfield  {journal} {\bibinfo  {journal} {IEEE Trans. Power App. Syst.}\ }\textbf {\bibinfo {volume} {PAS-89}},\ \bibinfo {pages} {2043} (\bibinfo {year} {1970})}\BibitemShut {NoStop}%
\bibitem [{\citenamefont {You}\ \emph {et~al.}(2017)\citenamefont {You}, \citenamefont {Zhang}, \citenamefont {Guo}, \citenamefont {Xu}, \citenamefont {Ma}, \citenamefont {Qin}, \citenamefont {Wen},\ and\ \citenamefont {Li}}]{7909196}%
  \BibitemOpen
  \bibfield  {author} {\bibinfo {author} {\bibfnamefont {H.}~\bibnamefont {You}}, \bibinfo {author} {\bibfnamefont {Q.}~\bibnamefont {Zhang}}, \bibinfo {author} {\bibfnamefont {C.}~\bibnamefont {Guo}}, \bibinfo {author} {\bibfnamefont {P.}~\bibnamefont {Xu}}, \bibinfo {author} {\bibfnamefont {J.}~\bibnamefont {Ma}}, \bibinfo {author} {\bibfnamefont {Y.}~\bibnamefont {Qin}}, \bibinfo {author} {\bibfnamefont {T.}~\bibnamefont {Wen}}, \ and\ \bibinfo {author} {\bibfnamefont {Y.}~\bibnamefont {Li}},\ }\href {\doibase 10.1109/TDEI.2017.006210} {\bibfield  {journal} {\bibinfo  {journal} {IEEE Trans. Dielectr. Electr. Insul.}\ }\textbf {\bibinfo {volume} {24}},\ \bibinfo {pages} {876} (\bibinfo {year} {2017})}\BibitemShut {NoStop}%
\bibitem [{\citenamefont {Cooke}\ \emph {et~al.}(1977)\citenamefont {Cooke}, \citenamefont {Wootton},\ and\ \citenamefont {Cookson}}]{21}%
  \BibitemOpen
  \bibfield  {author} {\bibinfo {author} {\bibfnamefont {C.}~\bibnamefont {Cooke}}, \bibinfo {author} {\bibfnamefont {R.}~\bibnamefont {Wootton}}, \ and\ \bibinfo {author} {\bibfnamefont {A.}~\bibnamefont {Cookson}},\ }\href {\doibase 10.1109/T-PAS.1977.32390} {\bibfield  {journal} {\bibinfo  {journal} {IEEE Trans. Power App. Syst.}\ }\textbf {\bibinfo {volume} {96}},\ \bibinfo {pages} {768} (\bibinfo {year} {1977})}\BibitemShut {NoStop}%
\bibitem [{\citenamefont {Endo}\ \emph {et~al.}(1980)\citenamefont {Endo}, \citenamefont {Kichikawa}, \citenamefont {Ishikawa},\ and\ \citenamefont {Ozawa}}]{23}%
  \BibitemOpen
  \bibfield  {author} {\bibinfo {author} {\bibfnamefont {F.}~\bibnamefont {Endo}}, \bibinfo {author} {\bibfnamefont {T.}~\bibnamefont {Kichikawa}}, \bibinfo {author} {\bibfnamefont {R.}~\bibnamefont {Ishikawa}}, \ and\ \bibinfo {author} {\bibfnamefont {J.}~\bibnamefont {Ozawa}},\ }\href {\doibase 10.1109/TPAS.1980.319713} {\bibfield  {journal} {\bibinfo  {journal} {IEEE Trans. Power App. Syst.}\ }\textbf {\bibinfo {volume} {PAS-99}},\ \bibinfo {pages} {847} (\bibinfo {year} {1980})}\BibitemShut {NoStop}%
\bibitem [{\citenamefont {Wu}\ \emph {et~al.}(2019{\natexlab{b}})\citenamefont {Wu}, \citenamefont {Zhang}, \citenamefont {Zhang}, \citenamefont {Guo}, \citenamefont {Du},\ and\ \citenamefont {Pang}}]{25}%
  \BibitemOpen
  \bibfield  {author} {\bibinfo {author} {\bibfnamefont {Z.}~\bibnamefont {Wu}}, \bibinfo {author} {\bibfnamefont {Q.}~\bibnamefont {Zhang}}, \bibinfo {author} {\bibfnamefont {L.}~\bibnamefont {Zhang}}, \bibinfo {author} {\bibfnamefont {C.}~\bibnamefont {Guo}}, \bibinfo {author} {\bibfnamefont {Q.}~\bibnamefont {Du}}, \ and\ \bibinfo {author} {\bibfnamefont {L.}~\bibnamefont {Pang}},\ }\href {\doibase 10.1088/1361-6595/ab32f4} {\bibfield  {journal} {\bibinfo  {journal} {Plasma Sources Sci. Technol.}\ }\textbf {\bibinfo {volume} {28}},\ \bibinfo {pages} {085018} (\bibinfo {year} {2019}{\natexlab{b}})}\BibitemShut {NoStop}%
\bibitem [{\citenamefont {Tian}\ \emph {et~al.}(2019)\citenamefont {Tian}, \citenamefont {Lin}, \citenamefont {Deng},\ and\ \citenamefont {Zhao}}]{Q01}%
  \BibitemOpen
  \bibfield  {author} {\bibinfo {author} {\bibfnamefont {Z.}~\bibnamefont {Tian}}, \bibinfo {author} {\bibfnamefont {H.}~\bibnamefont {Lin}}, \bibinfo {author} {\bibfnamefont {Y.}~\bibnamefont {Deng}}, \ and\ \bibinfo {author} {\bibfnamefont {H.}~\bibnamefont {Zhao}},\ }\href {\doibase 10.1063/1.5123183} {\bibfield  {journal} {\bibinfo  {journal} {Phys. Plasmas}\ }\textbf {\bibinfo {volume} {26}},\ \bibinfo {pages} {123517} (\bibinfo {year} {2019})}\BibitemShut {NoStop}%
\bibitem [{\citenamefont {Seeger}\ and\ \citenamefont {Clemen}(2013)}]{Seeger2}%
  \BibitemOpen
  \bibfield  {author} {\bibinfo {author} {\bibfnamefont {M.}~\bibnamefont {Seeger}}\ and\ \bibinfo {author} {\bibfnamefont {M.}~\bibnamefont {Clemen}},\ }\href {\doibase 10.1088/0022-3727/47/2/025202} {\bibfield  {journal} {\bibinfo  {journal} {J. Phys. D: Appl. Phys.}\ }\textbf {\bibinfo {volume} {47}},\ \bibinfo {pages} {025202} (\bibinfo {year} {2013})}\BibitemShut {NoStop}%
\bibitem [{\citenamefont {Wang}\ \emph {et~al.}(2022)\citenamefont {Wang}, \citenamefont {Sun},\ and\ \citenamefont {Teunissen}}]{wangzhen2022}%
  \BibitemOpen
  \bibfield  {author} {\bibinfo {author} {\bibfnamefont {Z.}~\bibnamefont {Wang}}, \bibinfo {author} {\bibfnamefont {A.}~\bibnamefont {Sun}}, \ and\ \bibinfo {author} {\bibfnamefont {J.}~\bibnamefont {Teunissen}},\ }\href {\doibase 10.1088/1361-6595/ac417b} {\bibfield  {journal} {\bibinfo  {journal} {Plasma Sources Sci. Technol.}\ }\textbf {\bibinfo {volume} {31}},\ \bibinfo {pages} {015012} (\bibinfo {year} {2022})}\BibitemShut {NoStop}%
\bibitem [{\citenamefont {Levko}\ and\ \citenamefont {Raja}(2023)}]{reaction4}%
  \BibitemOpen
  \bibfield  {author} {\bibinfo {author} {\bibfnamefont {D.}~\bibnamefont {Levko}}\ and\ \bibinfo {author} {\bibfnamefont {L.~L.}\ \bibnamefont {Raja}},\ }\href {\doibase 10.1063/5.0131780} {\bibfield  {journal} {\bibinfo  {journal} {J. Appl. Phys.}\ }\textbf {\bibinfo {volume} {133}},\ \bibinfo {pages} {053301} (\bibinfo {year} {2023})}\BibitemShut {NoStop}%
\bibitem [{\citenamefont {Nijdam}\ \emph {et~al.}(2020)\citenamefont {Nijdam}, \citenamefont {Teunissen},\ and\ \citenamefont {Ebert}}]{Nijdam_2020}%
  \BibitemOpen
  \bibfield  {author} {\bibinfo {author} {\bibfnamefont {S.}~\bibnamefont {Nijdam}}, \bibinfo {author} {\bibfnamefont {J.}~\bibnamefont {Teunissen}}, \ and\ \bibinfo {author} {\bibfnamefont {U.}~\bibnamefont {Ebert}},\ }\href {\doibase 10.1088/1361-6595/abaa05} {\bibfield  {journal} {\bibinfo  {journal} {Plasma Sources Sci. Technol.}\ }\textbf {\bibinfo {volume} {29}},\ \bibinfo {pages} {103001} (\bibinfo {year} {2020})}\BibitemShut {NoStop}%
\bibitem [{\citenamefont {Komuro}\ \emph {et~al.}(2018)\citenamefont {Komuro}, \citenamefont {Matsuyuki},\ and\ \citenamefont {Ando}}]{Komuro_2018}%
  \BibitemOpen
  \bibfield  {author} {\bibinfo {author} {\bibfnamefont {A.}~\bibnamefont {Komuro}}, \bibinfo {author} {\bibfnamefont {S.}~\bibnamefont {Matsuyuki}}, \ and\ \bibinfo {author} {\bibfnamefont {A.}~\bibnamefont {Ando}},\ }\href {\doibase 10.1088/1361-6595/aadf5c} {\bibfield  {journal} {\bibinfo  {journal} {Plasma Sources Sci. Technol.}\ }\textbf {\bibinfo {volume} {27}},\ \bibinfo {pages} {105001} (\bibinfo {year} {2018})}\BibitemShut {NoStop}%
\bibitem [{\citenamefont {Sigmond}(1984)}]{secondary1984}%
  \BibitemOpen
  \bibfield  {author} {\bibinfo {author} {\bibfnamefont {R.~S.}\ \bibnamefont {Sigmond}},\ }\href {\doibase 10.1063/1.334126} {\bibfield  {journal} {\bibinfo  {journal} {J. Appl. Phys.}\ }\textbf {\bibinfo {volume} {56}},\ \bibinfo {pages} {1355} (\bibinfo {year} {1984})}\BibitemShut {NoStop}%
\bibitem [{\citenamefont {Briels}\ \emph {et~al.}(2008)\citenamefont {Briels}, \citenamefont {Kos}, \citenamefont {Winands}, \citenamefont {van Veldhuizen},\ and\ \citenamefont {Ebert}}]{Briels_2008}%
  \BibitemOpen
  \bibfield  {author} {\bibinfo {author} {\bibfnamefont {T.~M.~P.}\ \bibnamefont {Briels}}, \bibinfo {author} {\bibfnamefont {J.}~\bibnamefont {Kos}}, \bibinfo {author} {\bibfnamefont {G.~J.~J.}\ \bibnamefont {Winands}}, \bibinfo {author} {\bibfnamefont {E.~M.}\ \bibnamefont {van Veldhuizen}}, \ and\ \bibinfo {author} {\bibfnamefont {U.}~\bibnamefont {Ebert}},\ }\href {\doibase 10.1088/0022-3727/41/23/234004} {\bibfield  {journal} {\bibinfo  {journal} {J. Phys. D: Appl. Phys.}\ }\textbf {\bibinfo {volume} {41}},\ \bibinfo {pages} {234004} (\bibinfo {year} {2008})}\BibitemShut {NoStop}%
\bibitem [{\citenamefont {Teunissen}\ and\ \citenamefont {Ebert}(2017)}]{Teunissen_2017}%
  \BibitemOpen
  \bibfield  {author} {\bibinfo {author} {\bibfnamefont {J.}~\bibnamefont {Teunissen}}\ and\ \bibinfo {author} {\bibfnamefont {U.}~\bibnamefont {Ebert}},\ }\href {\doibase 10.1088/1361-6463/aa8faf} {\bibfield  {journal} {\bibinfo  {journal} {J. Phys. D: Appl. Phys.}\ }\textbf {\bibinfo {volume} {50}},\ \bibinfo {pages} {474001} (\bibinfo {year} {2017})}\BibitemShut {NoStop}%
\end{thebibliography}%

\end{document}